\documentclass[%
 reprint,
 amsmath,amssymb,
 aps,
prb,
]{revtex4-2}

\usepackage{physics}
\usepackage{graphicx}
\usepackage{dcolumn}
\usepackage{bm}
\usepackage{hyperref}

\usepackage[normalem]{ulem} 

\hypersetup{citecolor=black,colorlinks=false,urlcolor=black}

\usepackage[utf8]{inputenc}
\usepackage[T1]{fontenc}
\usepackage{graphicx}
\usepackage{amsfonts, amsmath, amsthm, amssymb} 
\usepackage{braket}
\usepackage{float}
\usepackage{amsthm}
\usepackage{color}
\usepackage[all]{nowidow}
\usepackage[dvipsnames]{xcolor} 

\hypersetup{citecolor=black,colorlinks=false,urlcolor=black} 
\usepackage[english]{babel}

\usepackage{csquotes}

\usepackage{hyperref}
\newcommand{\appref}[1]{%
  \hyperref[#1]{App.~\ref*{#1}}%
}
\usepackage{xcolor}

\begin{document}

\preprint{APS/123-QED}

\title{Quantum Fisher information as a witness of non-Markovianity and criticality in the spin-boson model}

\author{D. Parlato$^{1,\dag}$}\author{G. Di Bello$^{1,3\dag}$}\author{F. Pavan$^{1}$}\author{G. De Filippis$^{2,3}$}\author{C. A. Perroni$^{2,3}$} 
\affiliation{$^{1}$Dip. di Fisica E. Pancini - Università di Napoli Federico II - I-80126 Napoli, Italy}
\affiliation{$^{2}$SPIN-CNR and Dip. di Fisica E. Pancini - Università di Napoli Federico II - I-80126 Napoli, Italy}
\affiliation{$^{3}$INFN, Sezione di Napoli - Complesso Universitario di Monte S. Angelo - I-80126 Napoli, Italy}
\affiliation{$^{\dag}$Both authors contributed equally to this work.}

\begin{abstract}
The quantum Fisher information, the quantum analogue of the classical Fisher information, is a central quantity in quantum metrology and quantum sensing due to its connection to parameter estimation and fidelity susceptibility. Using numerically exact methods applied to a paradigmatic open quantum system, the spin-boson model, we calculate both static and dynamical quantum Fisher information matrix elements with respect to spin-bath couplings and magnetic field strengths. As the spin-bath interaction increases, we first show that the coupling-coupling matrix elements relative to the ground state of the Hamiltonian are linked to the entanglement growth and signal the Berezinskii-Kosterlitz-Thouless quantum phase transition through their non-monotonic behavior. We also point out that the static quantum Fisher information exhibits a non-perturbative behavior in the zero-coupling limit, which we justify with an analytic argument. Furthermore, we demonstrate that the time-dependent matrix elements can reveal non-Markovian effects as well as the transition from the coherent to incoherent regime at the Toulouse point, remaining robust under pure dephasing noise. Non-monotonic signatures of the quantum Fisher information matrix reflect changes in quantum resources such as entanglement and coherence, quantify non-Markovian behavior, and enable criticality-enhanced quantum sensing, thereby shedding light on key features of open quantum systems.
\end{abstract}

\maketitle

\section{Introduction}
The quantum Fisher information (QFI) is a transversal quantity linked to various aspects of quantum physics, making it a versatile tool to characterize any system \cite{QFIM,Discountities_QF,Gammelmark_2014,Alipour_2014,Sidhu_2020,Faist_2023,QFI_rabi,QFI_multipartite}. 
First of all, the quantum Fisher information matrix (QFIM) satisfies the quantum Cramér-Rao bound \cite{Gammelmark_2014,Alipour_2014,Nielsen2013,Helstrom_1976,Holevo_2001}: the variance of a parameter estimate is {lower bounded} by the corresponding QFI, which depends only on the quantum state rather than on the specific measurement process. This makes the QFI a central concept to enhance parameter estimation in quantum systems. More precisely, 
given a vector $\vec{x}=(x_1,.,x_{\mu},.,x_{\nu},.,x_n)$ containing the relevant parameters of a system, and a density operator $\rho(\vec{x})$ dependent on these parameters, the QFIM is defined as \cite{QFIM}:
\begin{equation}F_{\mu\nu}=\frac{1}{2}\,\text{Tr}(\rho \{L_{\mu},L_{\nu}\}) \; ,\label{Cap1_EQ_QFIM_definition}\end{equation}
where $\{A,B\}$ denotes the anti-commutator of the operators $A$ and $B$. The operators $L_{\mu}$ and $L_{\nu}$ are the {symmetric logarithmic derivative} operators, implicitly defined as the solution to the equation (for the parameter $x_{\mu}$)
$\partial_{\mu} \rho=\frac{1}{2}(L_{\mu}\rho+\rho L_{\mu}) \; , \quad \partial_{\mu}\equiv\frac{\partial}{\partial x_{\mu}}\;$.
By using the spectral decomposition of the density operator, the QFIM becomes \cite{QFIM}:
\begin{equation}
F_{\mu\nu}=\sum_{i,j} \frac{2Re(\bra{\lambda_i}\partial_{\mu} \rho \ket{\lambda_j} \bra{\lambda_j}\partial_{\nu} \rho\ket{\lambda_i})}{\lambda_i+\lambda_j}\; ,\label{Cap1_calcolo_Fisher}
\end{equation}
where $\lambda_i$ and $\ket{\lambda_i}$ are the eigenvalues and eigenvectors of $\rho$, respectively. 

In addition to parameter estimation, the QFIM has deep significance in understanding the geometric structure of the Hilbert space, as it quantifies the distinguishability between quantum states corresponding to different parameter values.
The Bures distance can be defined as 
$D_B^2(\rho_1,\rho_2)=2(1-f(\rho_1,\rho_2))\; , \label{EQ_buresdistance}$
where $f(\rho_1,\rho_2)$ is Uhlmann's fidelity \cite{UHLMANN} between the quantum states $\rho_1$ and $\rho_2$.
Considering the distance between infinitesimally close states, one can define the Bures metric matrix $g_{\mu\nu}$.
It is found that the Bures metric is proportional to the QFIM:
$4g_{\mu\nu}=F_{\mu\nu}\;$ \cite{Discountities_QF,correspondence}. This relation establishes the QFIM as the metric tensor in the parameter space of quantum states, quantifying state distinguishability and sensitivity under parameter variations. This property makes it central for quantum sensing, as it indicates how to achieve maximum parameter sensitivity, which is relevant for the development of quantum detectors \cite{space,Giovannetti_2006,quantum_est_technology, T_th_2014}

Near a quantum phase transition (QPT) \cite{Sachdev}, the ground state is maximally sensitive to the variation of its driving parameter. Indeed, the QFI can also be employed to characterize QPTs along with other quantum-information-related quantities, such as entanglement and fidelity \cite{referee1,referee2,referee3,GU_2010}, and criticality-enhanced sensing \cite{Zhu_2023,L__2022,QFI_enh_sensing,QFI_enh_sensing_2,adani2024critical}. Although the QFI can be related to other quantities, it provides a complementary, explicitly metrological perspective in affording direct access to state sensitivity. Using the definition of the Bures distance, a strong connection is found between fidelity susceptibility and the QFIM:
\begin{equation}f(\rho(\vec{x}),\rho(\vec{x}+d\vec{x}))=1-\frac{1}{8}\sum_{\mu\nu}F_{\mu\nu}dx_{\mu}dx_{\nu}\;.
\label{Cap1_EQ_fid_QFI_link}\end{equation}
A diverging fidelity susceptibility has already been linked to a QPT \cite{GU_2010,Zhang_2010}, so that also the QFI tends to diverge in correspondence of such a transition. \\
The usefulness of QFI extends further beyond closed quantum systems, as it also offers insights into the behavior of the more realistic open systems \cite{breuer2002theory,Shaller,quantum_est_technology,Hao_2013,Wang_2015}. These systems typically exhibit relaxation, decoherence, and, with increasing spin-bath coupling, non-Markovian dynamics \cite{Breuer2,Vega2}, the signatures of which lie in the study of the QFI, particularly in the QFI flow, i.e., its time derivative \cite{QFI_non_markov,Lu_2010}. However, to the best of our knowledge, a complete understanding of the static and dynamical behavior of the QFIM is still lacking for open quantum systems, especially in the regime of strong coupling with the environment. In particular, no studies so far have systematically addressed the role of QFIM as an indicator of environment-induced QPTs, leaving this aspect unexplored.\\
In this work, through numerically exact methods, we analyze the QFIM of a paradigmatic open quantum system, the spin-boson model (SBM) \cite{Quantum_diss_syst,breuer2002theory,Shaller}, which describes the interaction of a spin one-half, representing a generic two-level system (TLS), with a bosonic bath of $N$ harmonic oscillators. This model, which includes amplitude damping processes, exhibits a Berezinskii-Kosterlitz-Thouless (BKT) QPT for strong spin-bath interaction \cite{De_Filippis_2020,De_Filippis_2023,Di_Bello_2024}. First of all, we show that the maximum of the QFIM of the Hamiltonian ground state (GS) turns out to be a marker of the related QPT, signaling criticality-enhanced sensing to very weak magnetic fields. Moving from the weak to intermediate coupling regime, the coupling-coupling QFIM element shows a decreasing monotonic behavior linked to increasing von Neumann entropy. We uncover a non-perturbative divergence of the QFI in the zero-coupling limit and validate it by benchmarking our numerical data against a perturbative analytic expansion. Furthermore, we demonstrate that the QFIM dynamics witness the coherent-incoherent transition and provide useful insights into the non-Markovian character of the system through the QFI flow. In particular, in the coherent regime, this flow exhibits characteristic oscillations, whose frequencies are different from those of spin observables. \\ 
We then generalize the model to provide a comprehensive description
of the noise mechanisms acting on the spin, including a pure dephasing term, against which the relevant signatures of QFIM remain robust.
These findings prove that non-monotonic features of the QFIM elucidate key properties of open quantum systems.

The paper is structured as follows. In Sec.~\ref{sec:modelmethods} we present the spin-boson model analyzed in this work, along with the analytical and numerical techniques employed. Our results are reported in Sec.~\ref{sec:results}, where we analyze both the static QFIM in Subsec.~\ref{subsec:staticQFIM} and its dynamical behavior in Subsec.~\ref{subsec:dynQFIM}. Finally, we conclude and discuss our findings, together with possible perspectives, in Sec.~\ref{sec:conclusions}. In the appendices, we provide additional details on , on the static behavior of different QFIM elements and the estimation of the critical coupling (\appref{app:statQFIM}), weak and zero coupling to the bath (\appref{app:weakcoupling}), on the static and dynamical features of the $z$ component of the Bloch vector (\appref{app:sigmaz}), on the dynamical QFIM and its relation to non-Markovianity (\appref{app:dynQFIM}) and on the Lindblad approximation (\appref{app:lindQFIM}).

\section{Model and Methods}\label{sec:modelmethods}

We focus on the zero-temperature regime and adopt units where $\hbar=1$.
The Hamiltonian of the generalized SBM is written as \cite{Quantum_diss_syst}:
\begin{equation}H_{SB}=H_{Q}+H_B+H_I\;, \label{spin-boson-hamiltonian}\end{equation}
where: $1)$ $H_Q=-\frac{\Delta}{2}\sigma_x-\frac{h}{2}\sigma_z$ is the spin one-half free Hamiltonian, with $\Delta$ (our energy unit) and $h$ (typically small) representing magnetic fields along the $x$ and $z$ axes, respectively; $2)$ $H_B=\sum_{i=1}^N\omega_i\bigg(b_i^\dagger b_i+\frac{1}{2}\bigg)$ is the bath Hamiltonian, a set of $N$ harmonic oscillators with frequencies $\omega_i$; $3)$ $H_I=\sum_{i=1}^N(\sigma_z\lambda_i^z+\sigma_x\lambda_i^x)(b_i+b_i^\dagger)$ is the qubit-bath interaction Hamiltonian, with couplings $\lambda_i^x$ and $\lambda_i^z$. Here, $\sigma_z$ and $\sigma_x$ are the Pauli matrices representing the TLS, and $b_i$ and $b_i^\dagger$ the annihilation and creation operators for bosonic modes of frequency $\omega_i$.
The couplings $\lambda_i^z$ are obtained by discretizing a continuous spectral density function $J(\omega)=\sum_i^N|\lambda_i^z|^2\delta(\omega-\omega_i)=\frac{\alpha}{2}\omega_c^{1-s}\omega^s\Theta(\omega_c-\omega)$, where $\omega_c$ is the cutoff frequency, that we set to $\omega_c=10\Delta$, $\alpha$ is the dimensionless spin-bath coupling associated with amplitude damping processes, $s$ is the Ohmicity parameter which we set to $s=1$ (Ohmic bath). The couplings $\lambda_i^x$ are sampled analogously, and we denote the dimensionless strength of pure dephasing processes by $\kappa$. \\ 
We will compute the Hamiltonian GS and dynamics of the QFIM elements by using analytical approximations as well as numerical techniques. Analytical treatments of the dynamics considered in this work rely on Lindblad's quantum master equation \cite{lindblad} (within this approximation, the static stationary QFIM elements are erroneously zero \appref{app:lindQFIM}). This approach is valid at weak couplings for Markovian dynamics, but it breaks down when the system-bath interaction strengthens and non-Markovian effects emerge \cite{breuer2002theory,Shaller}. \\
To access the strong coupling regime and capture non-Markovian effects, we employ tensor-network techniques based on matrix product state (MPS) representation, that yields numerically exact results for models mapped onto one-dimensional chains, using ITensor library \cite{fishman2022itensor}. The GS has been calculated using the density matrix renormalization group (DMRG) with a cutoff of $10^{-15}$ \cite{schollwock2011density}. The dynamics have been simulated starting from a spin-down state and the vacuum for the bath oscillators, using the $W^{II}$ algorithm \cite{paeckel2019time}, i.e. applying on the state a matrix product operator constructed from the second order expansion of the evolution operator. In order to get convergence in numerical simulations, we have used $N=600$ bosonic modes calculating the QFIM through Eq.\,\eqref{Cap1_calcolo_Fisher}.\\

\section{Results}\label{sec:results}
In the following, we present the results obtained by computing QFIM at equilibrium (DMRG) and out of equilibrium (time-dependent MPS simulations and the Lindblad equation).
\subsection{Static QFIM through DMRG}\label{subsec:staticQFIM}
\begin{figure}[htbp]

        \includegraphics[scale=0.22]{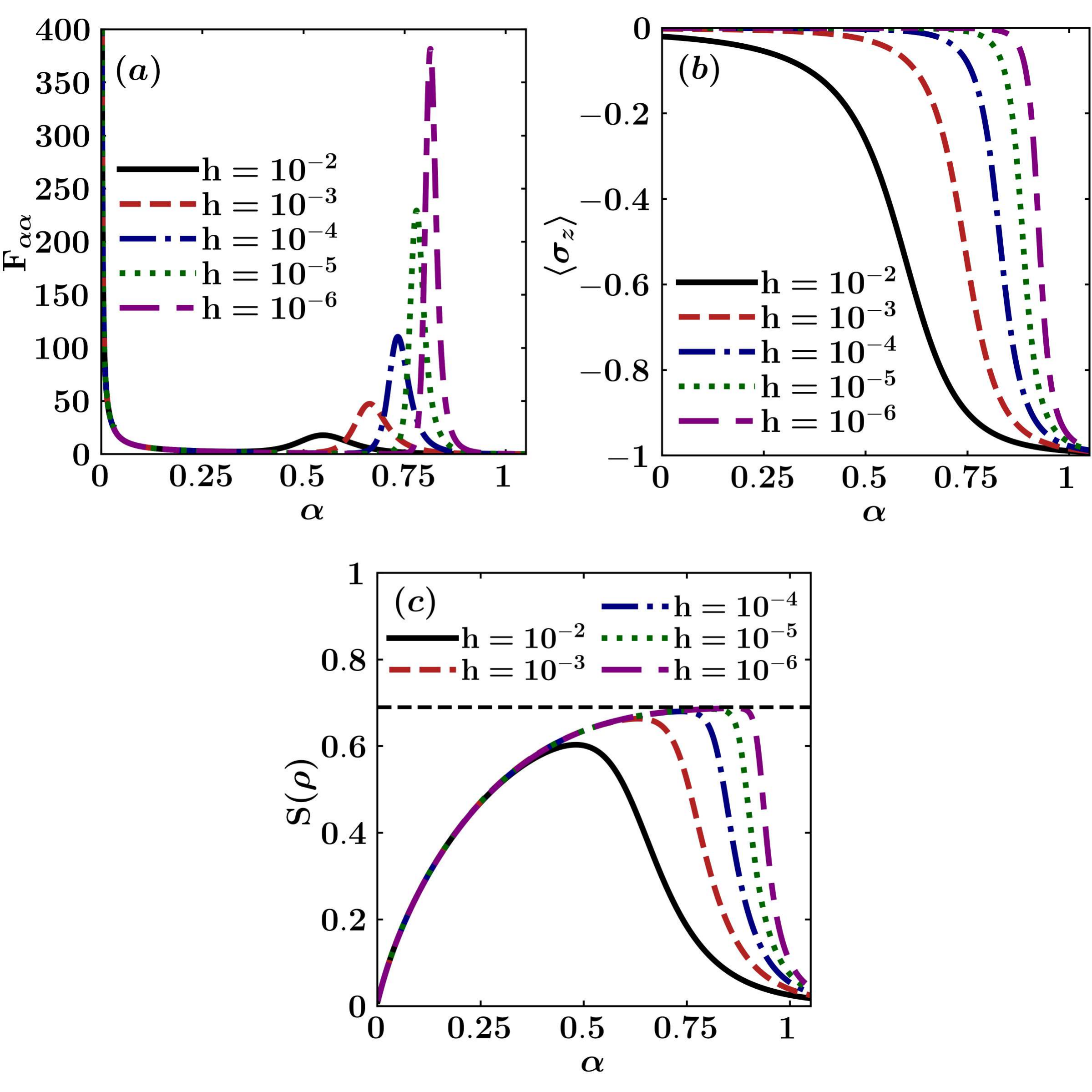}

    \caption{
    QFIM element $F_{\alpha\alpha}$ (a), $\langle\sigma_z\rangle$ (b), and von Neumann entropy (c) of the Hamiltonian GS as functions of spin-bath coupling $\alpha$ for several values of the magnetic field $h$ (in units of $\Delta$). 
    The dashed line in panel c indicates the maximum possible value $S_{max}=\ln2\approx0.69$ for a TLS. In all the plots, we set the spin-bath dephasing coupling $\kappa=0$.
    }
    \label{Static_QFI_new}
\end{figure}

We first focus on the QFIM calculated for the Hamiltonian GS in the absence of pure dephasing term ($\kappa=0$). We analyze the QFIM element $F_{\alpha\alpha}$ (see panel a of Fig.\,\ref{Static_QFI_new}), $F_{\alpha\Delta}$, and $F_{\Delta\Delta}$ (see panels a and b of Fig.\,\ref{Static_QFI} in \appref{app:statQFIM}) as functions of the spin-bath coupling $\alpha$ for small positive values of the magnetic field $h$. The limit $h\rightarrow 0^+$ is relevant due to the symmetry breaking that occurs in these systems. Actually, the SBM is known to exhibit a QPT \cite{Florens_2010,De_Filippis_2020,Zhang_2010,Tong_2011,Hur_2008} from a delocalized quantum to a localized classical spin state at the critical value $\alpha_c \approx 1+O(\Delta/\omega_c)$ ($\alpha_c \approx1.05$ for our choice of parameters).

For any small $h$, all three QFIM elements display a pronounced peak at strong coupling, meaning that the system becomes increasingly sensitive to variations in the coupling $\alpha$. With decreasing $h$, near criticality, the peak of the QFI $F_{\alpha\alpha}$ shifts towards larger values of $\alpha$ and becomes sharper and higher. Hence, for $h=0^+$, $F_{\alpha\alpha}$ is expected to diverge at the exact value $\alpha=\alpha_c$. As discussed in \appref{app:statQFIM}, the same behavior applies to the other two QFIM elements $F_{\alpha\Delta}$ and $F_{\Delta\Delta}$. Indeed, as the spin transitions from the quantum delocalized to the classical localized state, GSs corresponding to arbitrarily close values of $\alpha$ on opposite sides of the transition become nearly orthogonal, leading to a divergence of the QFIM. In \appref{app:statQFIM}, we detail the procedure used to extract the critical value $\alpha_c$ directly from QFIM elements and based on the position of their maxima. With this method we obtain a critical coupling $\alpha_c=1.03 \pm 0.03$, in excellent agreement with values reported in the literature \cite{De_Filippis_2020} (the extrapolations from $F_{\Delta\Delta}$ yield similar results). \\
Typically, the QPT of the SBM is characterized through spin observables like $\langle\sigma_z\rangle$ as $h$ decreases (see for example \cite{Hur_2008}). For this reason, in panel b of Fig.\,\ref{Static_QFI_new}, we report $\langle\sigma_z\rangle$ as a function of spin-bath coupling $\alpha$ for small positive of values $h$, showing perfect agreement with behaviors discussed in literature. Indeed, as $h$ decreases, 
the curves of $\langle\sigma_z\rangle$ steepen approaching the critical value of $\alpha$. Notably, the maxima of the QFIM correspond to the inflection points of the $\langle\sigma_z\rangle$, which, as discussed later, mark regions of high sensitivity not only for static but also for dynamical properties.\\
From the weak to intermediate coupling regime, $F_{\alpha\alpha}$ shows a monotonic decreasing behavior, which cannot be easily related to features of $\langle\sigma_z\rangle$.
As detailed in \appref{app:weakcoupling}, $F_{\alpha\alpha}$ diverges precisely as $1/\alpha$ in the ultra-weak coupling regime (see Fig.\,\ref{lowalpha}). This behavior is relevant since it implies very high state sensitivity for small $\alpha$, but its rapid decrease as $\alpha$ increases. 
Moreover, this feature recalls the connection of QFI with correlation functions \cite{QFIM,QFI_multipartite}: indeed, in related models with electron-oscillator interactions, the optical conductivity shows an inverse proportionality law with respect to the coupling strength \cite{De_Filippis_2014}. 
As shown in panel a of Fig.\,\ref{Static_QFI_new}, this $1/\alpha$ behavior persists up to intermediate coupling and, as discussed later, it remains robust under the introduction of the dephasing coupling $\kappa>0$. Perturbative analysis of Eq.\,\eqref{Cap1_calcolo_Fisher} clarifies this result. For $\alpha\rightarrow 0$, a first-order expansion gives $\langle\sigma_x\rangle\approx1+\alpha c$ with $c=-\bigg[\ln{\bigg(1+\frac{\omega_c}{\Delta}\bigg)-\frac{\omega_c}{\Delta+\omega_c}}\bigg]=-1.48$ for our choice of parameters. Furthermore, in the delocalized phase one finds $\langle\sigma_z\rangle=0$, and, due to the symmetry of the Hamiltonian, $\langle\sigma_y\rangle=0$. Substituting into Eq.\,\eqref{Cap1_calcolo_Fisher} yields $F_{\alpha\alpha}=c^2/[1-(1+c\alpha)^2]\sim -c/(2\alpha)$, confirming the non-perturbative divergence. See the Appendices \appref{app:weakcoupling} and \appref{app:statQFIM} for the full derivation and for the different behavior of $F_{\alpha \Delta}$ and $F_{\Delta\Delta}$.

Finally, in the weak to intermediate spin-bath coupling, the decreasing behavior of $F_{\alpha\alpha}$ can be linked to an increasing GS von Neumann entropy $S(\rho)$ of the spin one-half, reported in panel c of Fig.\,\ref{Static_QFI_new}. As the spin entangles more with the bath, its sensitivity decreases. In particular, at fixed $h$, the monotonic decrease of the QFI stops when the von Neumann entropy reaches approximately the maximum value. 

\begin{figure}[htbp]
        \includegraphics[scale=0.25]{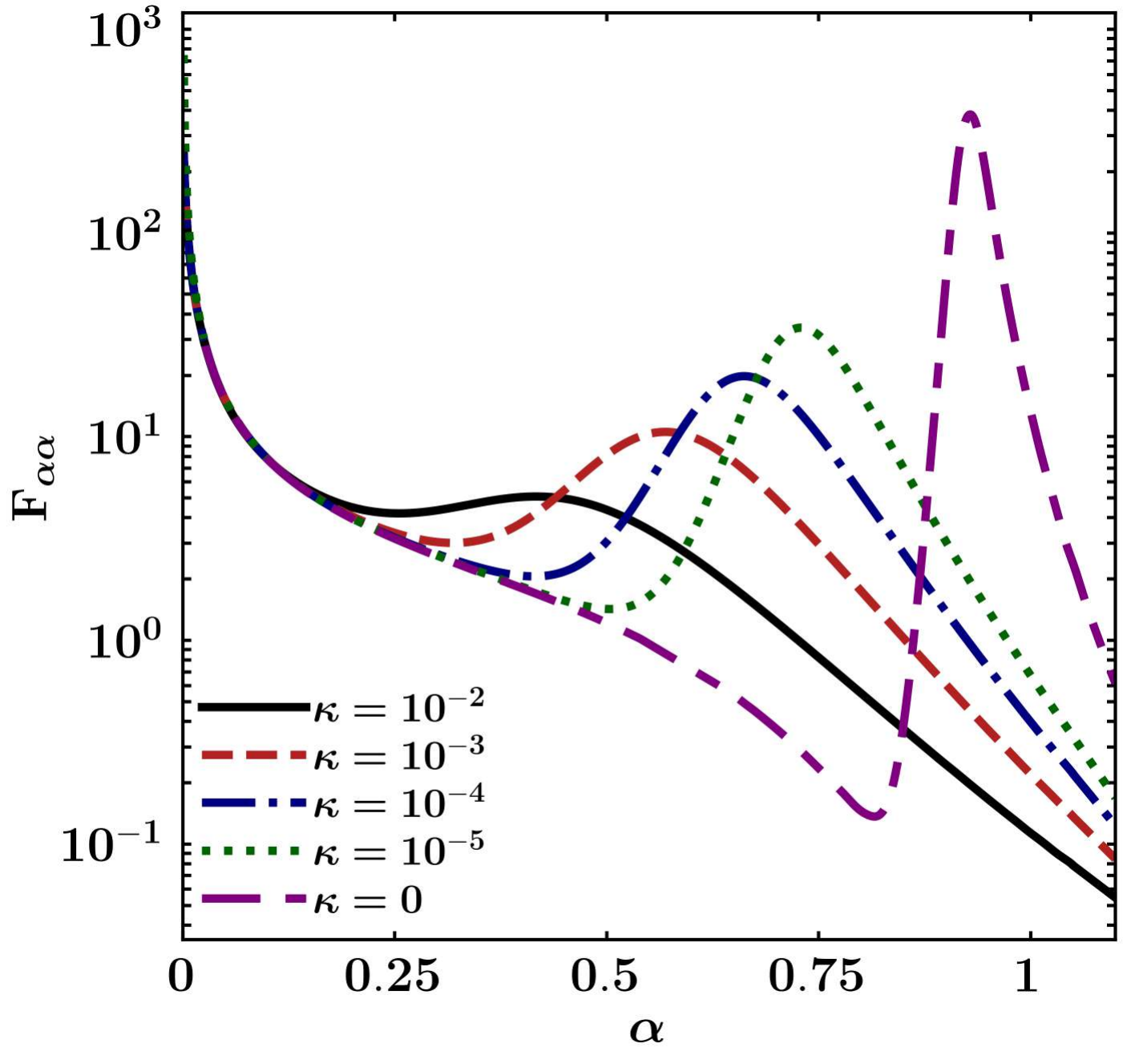}

    \caption{GS QFIM element $F_{\alpha\alpha}$ (in logarithmic scale) of the generalized SBM as a function of the spin-bath coupling $\alpha$, for $h=10^{-6}$ (in units of $\Delta$) and different values of the dephasing coupling $\kappa$. The $\kappa=0$ curve corresponds to the $h=10^{-6}$ case in Fig.\,\ref{Static_QFI_new} (a): its minimum becomes visible only in logarithmic scale.}
    \label{F_gamma}
\end{figure}

Up to this point, we have restricted our analysis to the case $\kappa=0$ (zero pure dephasing term). We now examine the QFIM element $F_{\alpha\alpha}$ of the GS of the full Hamiltonian in Eq.\,\eqref{spin-boson-hamiltonian}. As illustrated in Fig.\,\ref{F_gamma}, under weak dephasing, the system retains high sensitivity in the vicinity of the critical point for strong $\alpha$. One can interpret these results making a comparison with the plots of panel a of Fig.\,\ref{Static_QFI_new}, considering that the spin-bath coupling $\kappa$ in the Hamiltonian of Eq.\,\eqref{spin-boson-hamiltonian} effectively acts as an additional dynamical external field along $\sigma_z$, hence reducing the height of the maximum as $\kappa$ increases. These results are further supported by the behavior of $\langle\sigma_z\rangle$ reported in Fig.\,\ref{sigma_z_static} of the \appref{app:sigmaz} for the same model parameters and are similar to those obtained for QFIM elements $F_{\Delta\Delta}$, $F_{\alpha\Delta}$ (see \appref{app:statQFIM}). Importantly, $F_{\alpha\alpha}$ continues to diverge as $1/\alpha$ in the regime of very weak spin-bath coupling $\alpha$, demonstrating that the non-monotonic profile of the QFI as a function of $\alpha$ persists even with small but finite dephasing ($\kappa>0$). 

\subsection{Dynamical QFIM and its flux to witness non-Markovianity}\label{subsec:dynQFIM}
\begin{figure}[htbp]

        \includegraphics[scale=0.2]{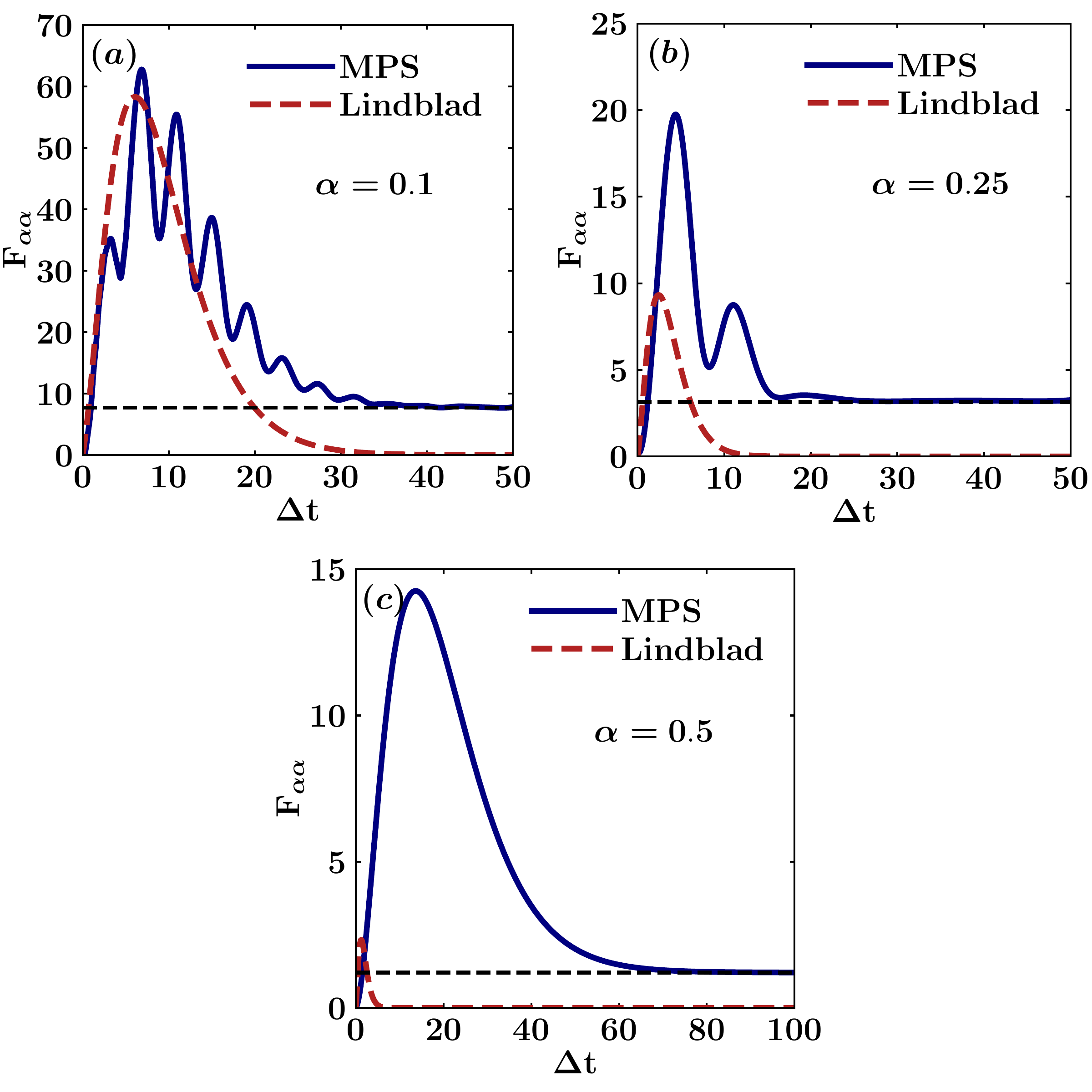}

    \caption{QFIM element $F_{\alpha\alpha}$, calculated with methods based on MPS and Lindblad master equation, as a function of time for spin-bath coupling $\alpha=0.1$ (a), $\alpha=0.25$ (b) and $\alpha=0.5$ (c), at the spin-bath coupling $\kappa=0$. The dashed horizontal line is the ground state QFI value calculated with the DMRG method: $F_{\alpha\alpha}^{\alpha=0.1}=7.7$, $F_{\alpha\alpha}^{\alpha=0.25}=3.15$, and $F_{\alpha\alpha}^{\alpha=0.5}=1.21$.}
    \label{F_alpha_tempo}
\end{figure}
We now turn to the analysis of the information encoded in the time evolution of the QFIM. In particular, we focus on the evolution of coupling-coupling element $F_{\alpha\alpha}(t)$, computed using both the numerically exact MPS approach \cite{Di_Bello2_2024} and the Markovian Lindblad master equation, the latter being valid only in the weak spin-bath coupling regime \cite{Saleem}. Within the framework of Lindblad's approximation, the analytical form of the density operator $\rho(t)$ allows us to derive a closed-form expression for the QFIM, as detailed in \appref{app:lindQFIM}. In Fig.\,\ref{F_alpha_tempo}, we show the QFI element $F_{\alpha\alpha}$ as a function of time for the spin-bath amplitude couplings $\alpha=0.1, 0.25, 0.5$ and dephasing coupling $\kappa=0$ in the case of $h=0$ (in this regime of parameters, very small values of $h$ introduce only negligible corrections). In the \appref{app:dynQFIM}, we analyze also the role played by finite values of dephasing coupling $\kappa>0$.\\
The dynamics of the SBM exhibit a coherent-incoherent transition at the Toulouse point $\alpha=0.5$: for $\alpha<0.5$, the system displays coherent oscillations, whereas for $\alpha>0.5$ the behavior becomes incoherent \cite{kennes2013oscillatory}. This can be seen in the dynamics of $\langle\sigma_z\rangle$ that shows these two distinct behaviors \cite{Hur_2008}, as shown in Fig.\,\ref{sigma_z_tempo} of the \appref{app:sigmaz} for the same parameters as in Fig.\,\ref{F_alpha_tempo}. \\
We first examine the coherent regime (see panels a and b of Fig.\,\ref{F_alpha_tempo}). For $\alpha=0.1$, MPS and Lindblad predictions of QFI $F_{\alpha\alpha}$ are comparable only at short times, differing significantly as the stationary state is approached. Additional results reported in the \appref{app:weakcoupling} show that even for $\alpha=0.01$, where the two approaches nearly coincide initially, they still deviate at long times (see Fig.\,\ref{F_alpha_0.01}). Actually, the behavior of $F_{\alpha\alpha}(t)$ within Lindblad approximation can be separated into two regimes: an initial transient characterized by quadratic growth, $F_{\alpha\alpha}(t) \sim t^2$, followed by a rapid exponential decay towards zero \cite{Saleem}. Indeed, the Lindblad QFI always vanishes at long times, since the corresponding steady state (the Gibbs zero temperature state) is independent of $\alpha$.
In contrast, the MPS QFI dynamics exhibit pronounced oscillations modulated by an exponentially decaying envelope. These oscillations progressively damp out, and the QFI converges to a finite stationary value, in excellent agreement with the DMRG GS results and, consequently, strongly dependent on $\alpha$. As the coupling strength $\alpha$ increases, both the frequency of oscillations and the peak amplitude of the QFI decrease, indicating a reduction in the system’s sensitivity. In this coherent regime, the QFIM flow \cite{Lu_2010,QFIM}, defined as $\partial_t F_{\mu,\nu}$, can be used to characterize non-Markovian effects after the initial transient. As detailed in \appref{app:dynQFIM}, starting from the maximum of the respective curves, the MPS solution exhibits alternating positive and negative flow, evidencing continuous information backflow between system and bath (see Fig.\,\ref{FF}). By contrast, the Lindblad flow remains strictly negative, indicating a purely outward flow of information.\\ 
In the coherent regime, the QFI dynamics are correlated with those of the spin observables. From the comparison with results of $\langle\sigma_z\rangle(t)$ reported in \appref{app:sigmaz} in Fig.\,\ref{sigma_z_tempo}, we find that each local maximum of $F_{\alpha\alpha}$ corresponds approximately to an extremum of $\langle \sigma_z\rangle (t)$. These points represent regions of high sensitivity to variations in $\alpha$, and therefore coincide with the peaks of the QFI. This interpretation is further supported by the Fourier analysis presented in Fig.\,\ref{FF} of the \appref{app:dynQFIM}: the dominant frequency component of $F_{\alpha\alpha}(t)$ is exactly twice that of $\langle\sigma_z\rangle(t)$, a property which remains robust upon pure dephasing noise. \\
Finally, as shown in panel c of Fig.\,\ref{F_alpha_tempo}, for $\alpha=0.5$ the QFI witnesses the transition to the incoherent regime: in fact oscillations vanish. Apart from an initial transient in which it reaches the maximum, it monotonically relaxes to the stationary DMRG value.

\section{Conclusions}\label{sec:conclusions}
We have analyzed static and dynamical QFIM in a generalized SBM, a paradigmatic prototype of open quantum systems. 
The QFIM element $F_{\alpha\alpha}$ exhibits, in the limit of zero spin-bath coupling, a divergent behavior, verified by analytic calculations, while, into the intermediate coupling regime, it is characterized by a monotonic decrease, correlated with an increase in the von Neumann entropy. 

In the strong coupling regime, the static QFIM reveals clear signatures of criticality-enhanced sensing near the BKT QPT, allowing to identify the values of the critical coupling $\alpha_c$.

In the coherent regime, the dynamical QFIM displays a characteristic oscillatory behavior whose dominant frequency is exactly twice that of spin observables, establishing a direct connection between QFI flow and non-Markovian effects.
Moreover, the dynamical QFIM reflects the coherent-to-incoherent transition that occurs as the spin interaction with the environment increases. Static and dynamic findings remain robust under pure dephasing noise proving that non-monotonic signatures of the QFIM provide powerful probes for identifying criticality, quantifying non-Markovian behavior, and understanding key phenomena in open quantum systems. 

\section*{Acknowledgements}
G.D.F. acknowledges financial support from PNRR MUR Project No. PE0000023-NQSTI. G.D.B. and C.A.P. acknowledge funding from IQARO (Spin-orbitronic Quantum Bits in Reconfigurable 2DOxides) project of the European Union’s Horizon Europe research and innovation programme under grant agreement n. 101115190. G.D.B., G.D.F. and C.A.P. acknowledge funding from the PRIN 2022 project 2022FLSPAJ ``Taming Noisy Quantum Dynamics'' (TANQU). C.A.P. acknowledges funding from the PRIN 2022 PNRR project P2022SB73K ``Superconductivity in KTaO3 Oxide-2DEG NAnodevices for Topological quantum Applications'' (SONATA) financed by the European Union - Next Generation EU. 
 
\bibliography{main}

\appendix

\section{Additional features of static QFIM elements and estimation of \texorpdfstring{$\alpha_c$}{alpha c}}\label{app:statQFIM}
\begin{figure}[htbp]
        \includegraphics[scale=0.28]{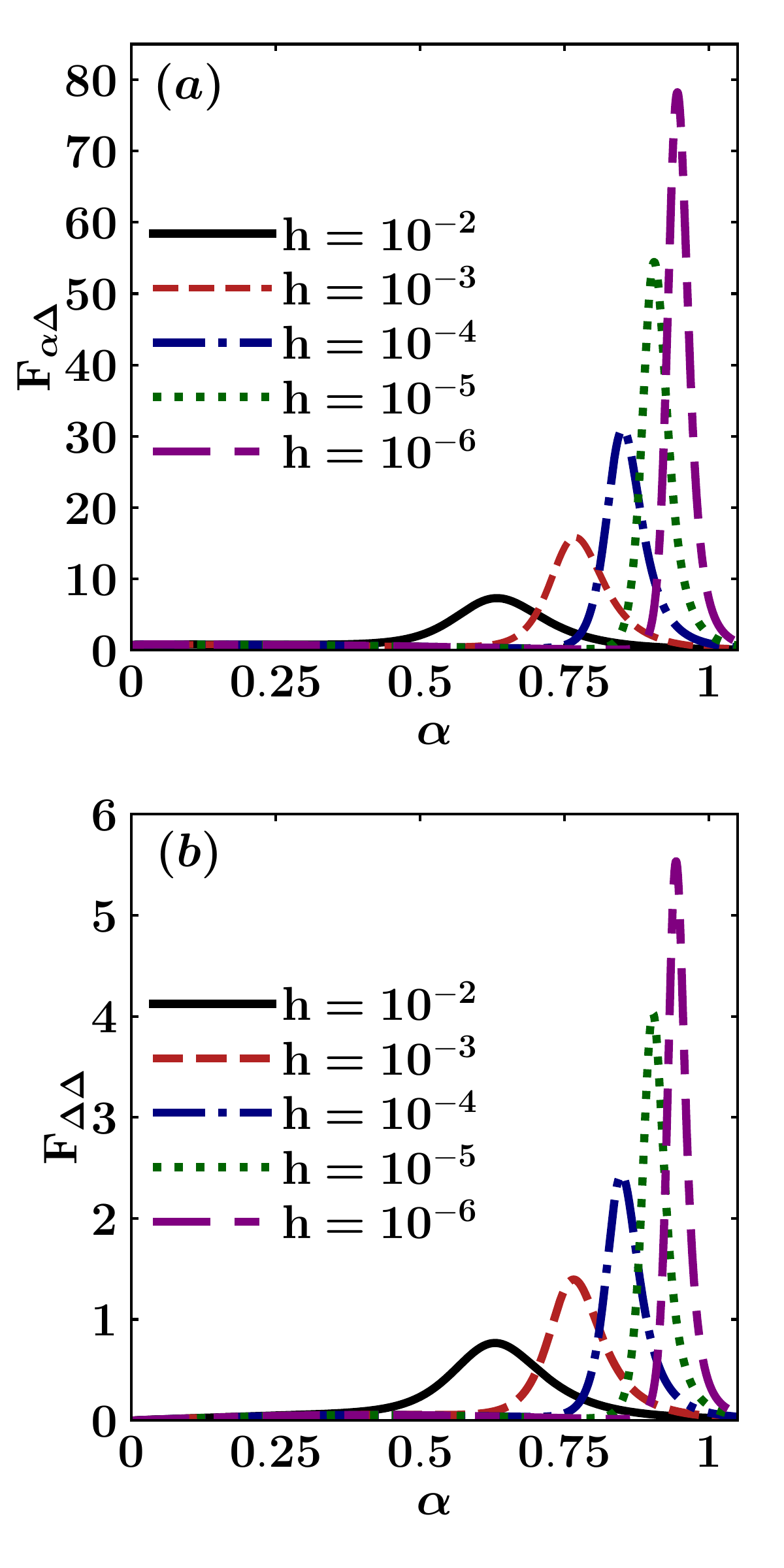}

    \caption{QFIM elements $F_{\alpha\Delta}$ (in units of $1/\Delta$) (a), and $F_{\Delta\Delta}$ (in units of $1/\Delta^2$) (b) as functions of spin-bath coupling $\alpha$ for various values of $h$ (in units of $\Delta$) at $\kappa=0$.}
    \label{Static_QFI}
\end{figure}

In Fig.\,\ref{Static_QFI}, we report the static QFIM elements $F_{\alpha\Delta}$ (a) and $F_{\Delta\Delta}$ (b) as functions of $\alpha$ for different small values of the magnetic field $h$. As discussed in the main text, density matrix renormalization group (DMRG) is used to compute these results \cite{schollwock2011density,nocera2}, and simple arguments based on perturbative estimates and dimensional analysis explain the reason why, as $\alpha\rightarrow0$, $F_{\alpha\Delta}$ is finite ($F_{\Delta\alpha}\rightarrow0.8$ in units of $1/\Delta$) and $F_{\Delta\Delta}$ vanishes.
Actually, the ground state (GS) sensitivity to a variation of $\Delta$ vanishes because the bath renormalization effect is negligible, hence for an infinitesimal $\Delta$ variation, the two GSs are almost identical. \\

For strong spin-bath coupling, $F_{\alpha\Delta}$ and $F_{\Delta\Delta}$ show the same peak positions as $F_{\alpha\alpha}$ in the main text.
The peak in $F_{\alpha\Delta}$ can be explained by the fact that, similarly to $F_{\alpha\alpha}$, it also quantifies the sensitivity to $\alpha$, although with a smaller magnitude. The peak in $F_{\Delta\Delta}$ can be understood because of the shift in the critical coupling value (at $h=0^+$) caused by a variation of the bare $\Delta$. Indeed, since $\alpha_c\approx1+O(\Delta/\omega_c)$ \cite{Hur_2008}, at the critical point, two states differing for a small variation of $\Delta$ are subject to two different critical couplings. This implies that one state is in a delocalized state and the other in a localized one. The difference between the two situations is that, in this case, the maximum is related to the variation of $\alpha_c$ caused by the change in $\Delta$, while for $F_{\alpha\alpha}$ the critical $\alpha_c$ is unchanged. This clearly implies that the height of the $F_{\Delta\Delta}$ peak is quite reduced.

\begin{figure}[htbp]
        \includegraphics[scale=0.28]{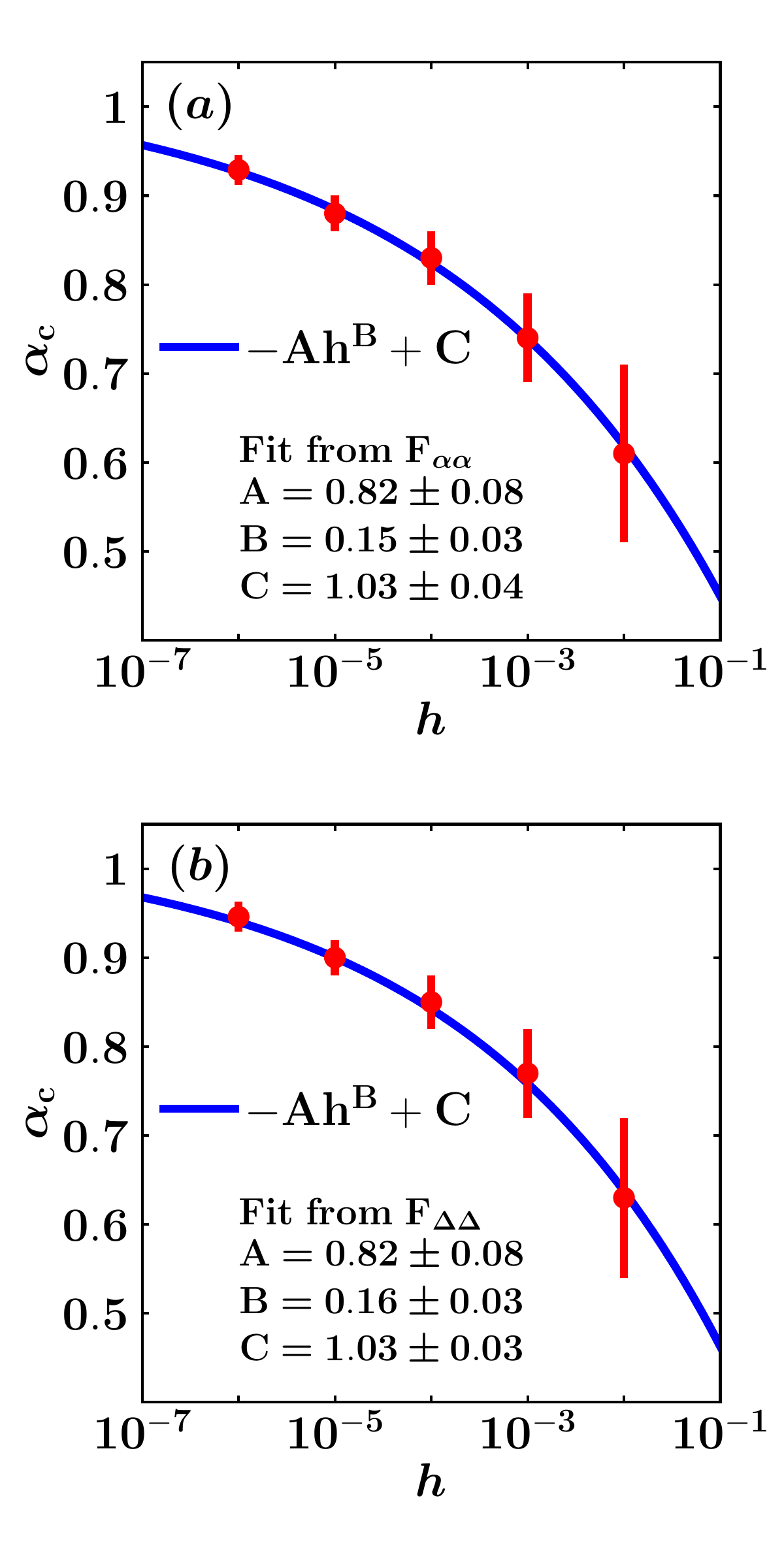}

    \caption{Extrapolation of the critical coupling $\alpha_c$ from the behavior of the QFI $F_{\alpha\alpha}$ (a) and $F_{\Delta\Delta}$ (b) as a function of the magnetic field $h$ (in units of $\Delta$). We also show the resulting power-law fit in Eq.\,\eqref{Fit}.}
    \label{extrapolation}
\end{figure}

As mentioned in the main text, we obtain excellent estimates of the QPT critical coupling $\alpha_c$ directly from QFIM elements. To this aim, we track the position of the maximum of QFIM element for each value of $h$, taking the full width at half maximum of each peak as the associated uncertainty. The data points are then fitted with the following power-law function:
\begin{equation}
\alpha(h)=-Ah^B+C\;,
\label{Fit}
\end{equation}
where $C=\alpha_c=\alpha(h=0^+)$ represents the value of the critical coupling, obtained in the limit of zero field. \\
In Fig.\,\ref{extrapolation}, we present the estimated $\alpha_c$ along with the scaling law of the critical coupling and the values obtained by evaluating $h\rightarrow0^+$.
Using this method, we obtain a critical coupling for $F_{\alpha\alpha}$ of $\alpha_c = 1.03 \pm 0.03$, which is consistent with the value reported in the literature \cite{De_Filippis_2020} (see panel a). The extrapolations from $F_{\Delta\Delta}$ (b) yield similar consistent results.

\section{Weak and zero coupling regime: data fitting and analytic derivation of (non-perturbative) divergent QFI behavior}\label{app:weakcoupling}
Apart from the regime of strong spin-bath coupling near the QPT, the weak coupling regime is also of experimental interest. For this reason, we analyze the static and dynamical behavior of the QFI for small values of $\alpha$. 

\begin{figure}[htbp]
    \includegraphics[scale=0.25]{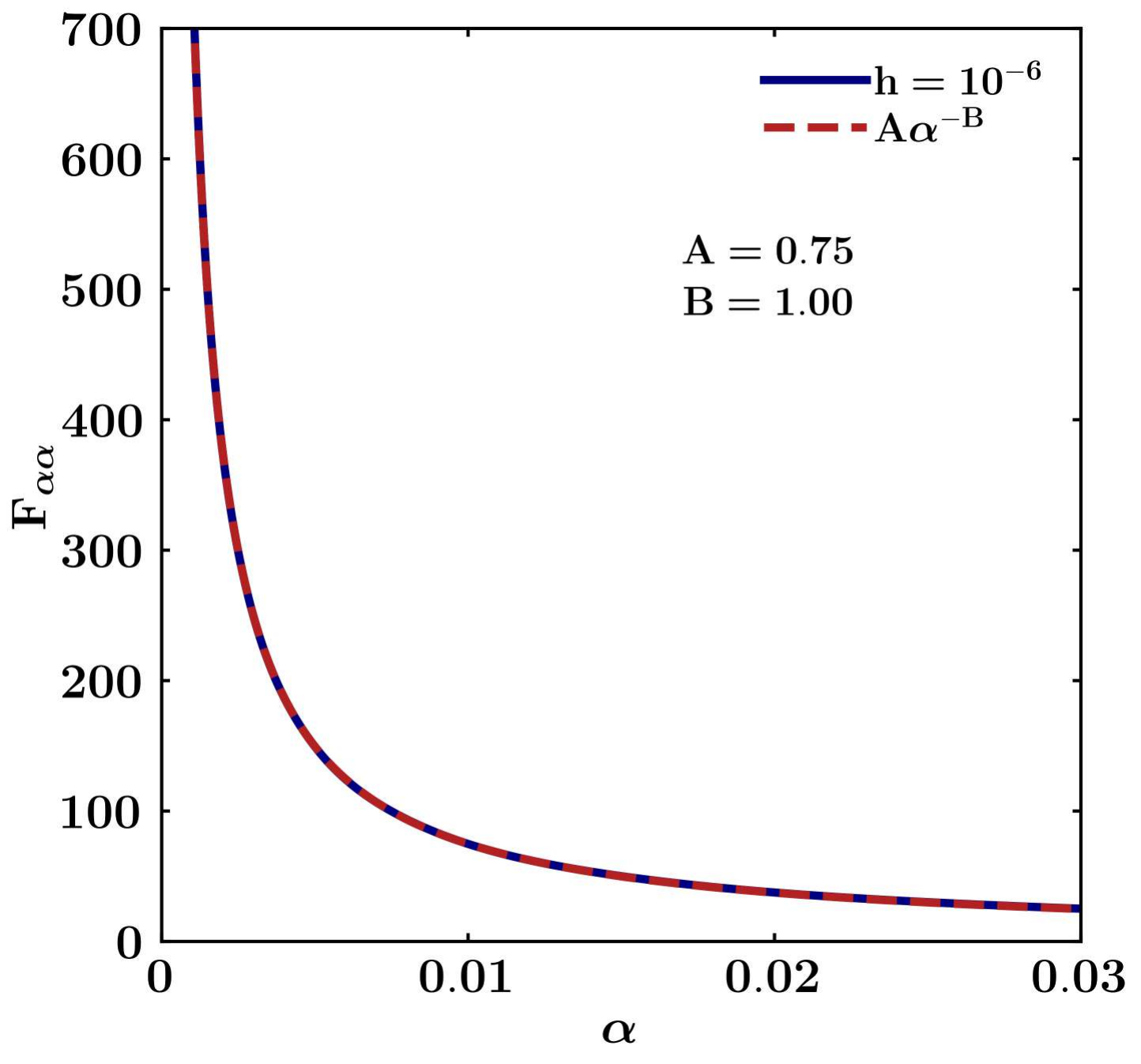}
       
    \caption{Zoomed-in view of the GS QFIM element $F_{\alpha\alpha}$ perfectly matching the power law in Eq.\,\eqref{Fit} in \appref{app:statQFIM} for $\alpha\in[0,0.03]$.}
    \label{lowalpha}
\end{figure}

\begin{figure}[htbp]
    \includegraphics[scale=0.25]{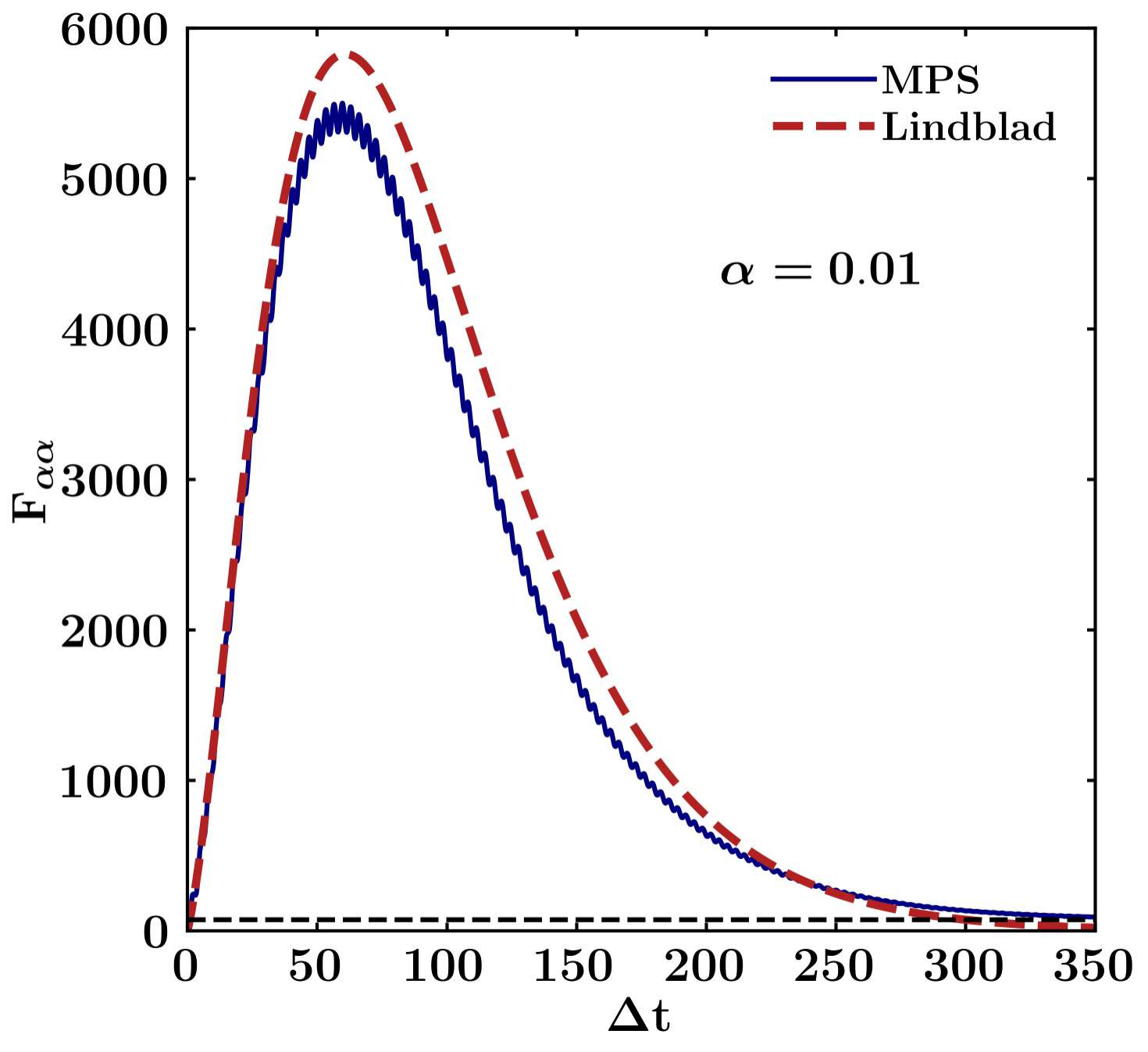}
    
    \caption{QFIM element $F_{\alpha\alpha}$ as a function of time for spin-bath coupling $\alpha=0.01$, $h=\kappa=0$: comparison of the results obtained with the MPS technique and the Lindblad method.}
    \label{F_alpha_0.01}
\end{figure}
 
As shown in panel a of Fig.\,\ref{Static_QFI_new} of the main text, $F_{\alpha\alpha}$ diverges as $\alpha\rightarrow0$. This behavior can be captured using a power-law fit of the form in Eq.\,\eqref{Fit} in \appref{app:statQFIM} with $C=0$, resulting in the scaling $F_{\alpha\alpha}\propto\alpha^{-1}$ for vanishing $\alpha$. 
In Fig.\,\ref{lowalpha}, we plot $F_{\alpha\alpha}$ in the range $\alpha\in[0,0.03]$ along with the resulting fit, showing an excellent agreement. This clearly indicates that the state sensitivity rapidly decreases with increasing $\alpha$, even for small values of $\alpha$, in a non-perturbative way, as we will demonstrate in the following. \\ 
We have also simulated the dynamics of the QFIM for $\alpha=0.01$ to study its behavior in the very weak coupling regime. As shown in Fig.\,\ref{F_alpha_0.01}, for weaker interactions, the amplitude of the oscillations is small compared to the QFI value. The coarse-graining approximation used in deriving the Lindblad quantum master equation is evident in this regime: Lindblad curve approaches the exact MPS result, averaging out the oscillations. Thus, as $\alpha\rightarrow0$, the MPS curve basically smooths out and coincides with the Lindblad one. In conclusion, for vanishing $\alpha$, the non-Markovian aspects of the dynamics disappear, as expected.

\begin{figure}[htbp]
    \includegraphics[scale=0.25]{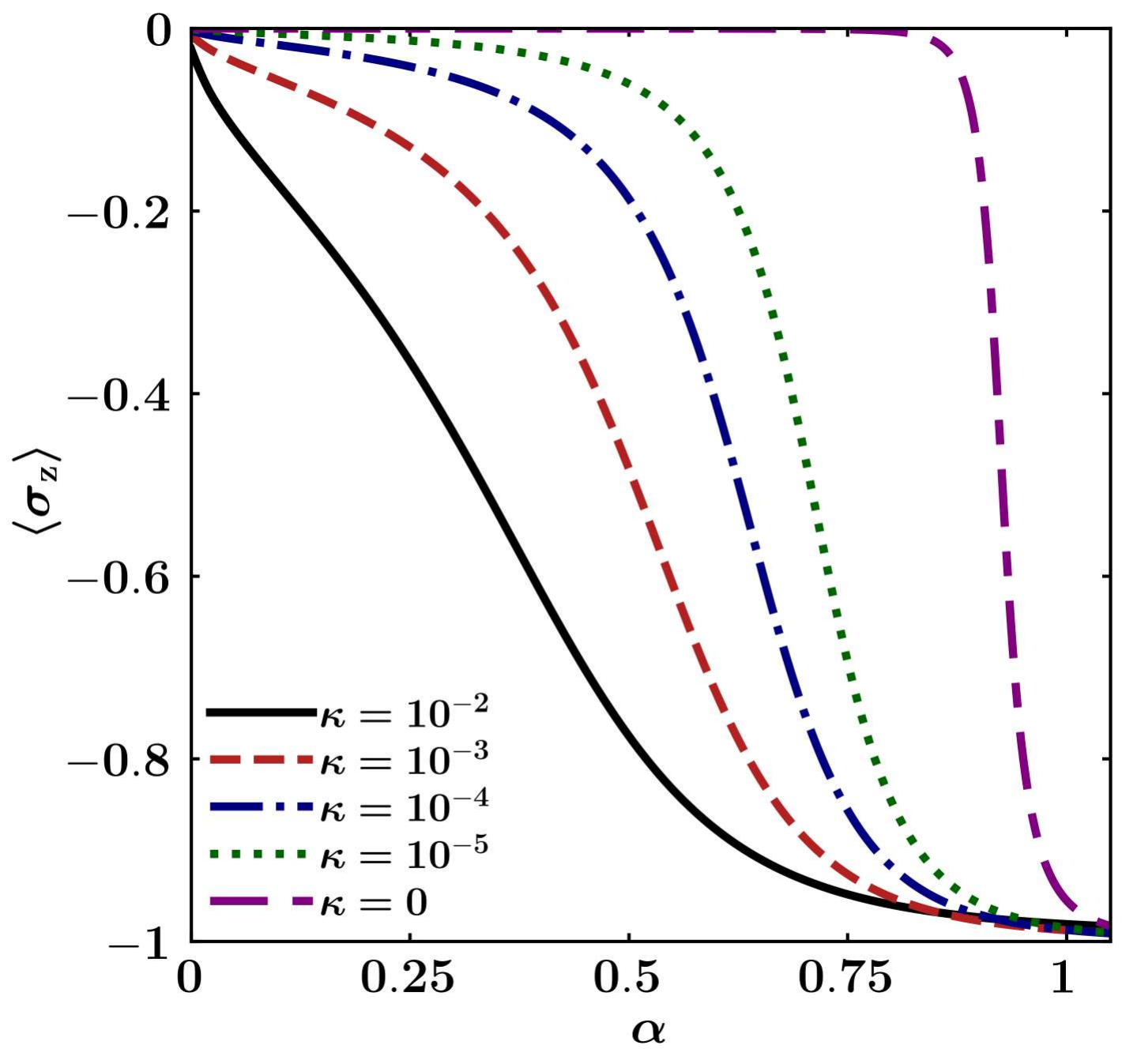}
    
    \caption{GS $\langle\sigma_z\rangle$ as a function of spin-bath coupling $\alpha$ for several values of spin-bath coupling $\kappa$ at magnetic field $h=10^{-6}$ (in units of $\Delta$).}
    \label{sigma_z_static}
\end{figure}

\begin{figure}[htbp]

    \includegraphics[scale=0.2]{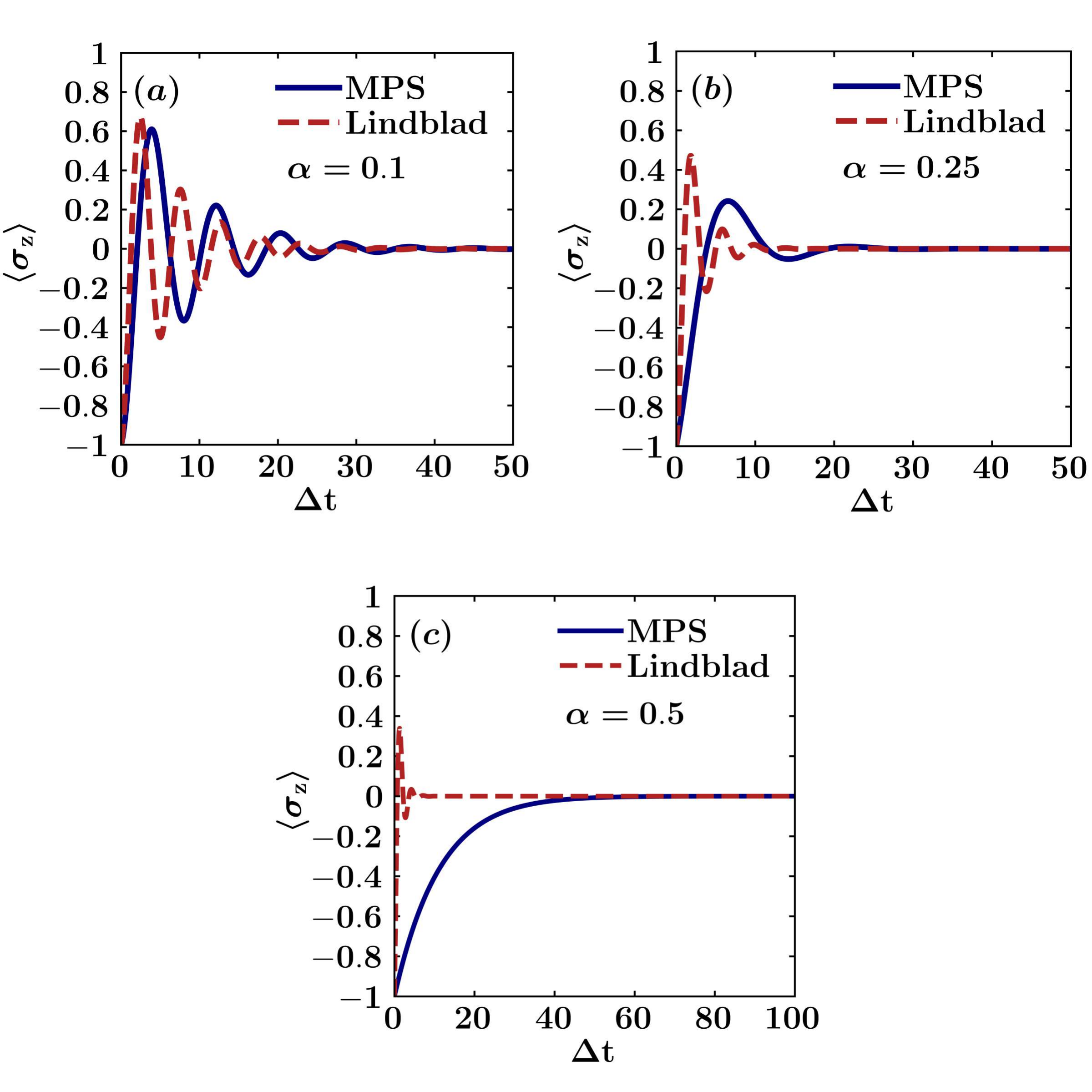}
    
    \caption{GS $\langle\sigma_z\rangle$ as a function of time for the spin-bath couplings $\alpha=0.1$ (a), $\alpha=0.25$ (b) and $\alpha=0.5$ (c), $h=0$ and spin-bath coupling $\kappa=0$.}
    \label{sigma_z_tempo}
\end{figure}

Now we provide an analytical explanation for this power-law scaling. First, we calculate the low-coupling behavior of the Bloch vector $\vec{r}=(\langle\sigma_x\rangle,\langle\sigma_y\rangle,\langle\sigma_z\rangle)$. Starting from $\langle\sigma_x\rangle$, a perturbative calculation with respect to $\alpha$ can be performed using the imaginary-time expansion \cite{mahan2013many}; to second order we obtain
\begin{equation}\langle \sigma_x\rangle=\frac{\langle\sigma_x\rangle_{H_0}+\int_0^\beta d\tau_1\int_0^{\tau_1} d\tau_2 \;[\langle H_I^{(0)}(\tau_1)H_I^{(0)}(\tau_2)\sigma_x\rangle]}{1+\int_0^\beta d\tau_1\int_0^{\tau_1} d\tau_2 \;[\langle H_I^{(0)}(\tau_1)H_I^{(0)}(\tau_2)\rangle]}\;,\end{equation} 
where $\langle A\rangle_{H_0}=Tr(\frac{e^{-\beta H_0}}{Z}A)$ denotes the average over the thermal state of the unperturbed Hamiltonian $H_0$, and $H_I^{(0)}(\tau)=e^{H_0\tau}H_Ie^{-H_0\tau}$ is the interaction Hamiltonian in the imaginary-time picture. After evaluating the integrals, the leading term at large $\beta$ is:
\begin{equation}\langle \sigma_x\rangle\approx\frac{1+\alpha(\beta A+c)}{1+\alpha\beta A}\;,\end{equation}
where $A=\int_0^{+\infty}d\omega \;[J(\omega)/\alpha(\Delta+\omega)]$ and $c=-\int_0^{+\infty}d\omega \;[J(\omega)/\alpha(\Delta+\omega)^2]$. We first take the small-coupling limit, 
therefore, using the previous spectral density function $J(\omega)=\frac{\alpha}{2}\omega\Theta(\omega_c-\omega)$, we get:
\begin{equation}\langle\sigma_x\rangle\approx\langle\sigma_x\rangle_{H_0}+\alpha c\;\end{equation}
with 
\begin{equation}\begin{array}{c}\langle\sigma_x\rangle_{H_0}=1,\\
c=-\bigg[\ln{\bigg(1+\frac{\omega_c}{\Delta}\bigg)}-\frac{\omega_c}{\omega_c+\Delta}\bigg]=-1.48,\end{array}\end{equation}
for our choice of parameters. This value of $c$ is also confirmed by a linear fit to the numerical data. The behavior of the other two Pauli matrices is trivial. Actually, because $H$ is independent of $\sigma_y$, we find that $\langle\sigma_y\rangle=0$ for any $\alpha$; and in the delocalized phase ($\alpha\rightarrow0$ limit considered here) we also have $\langle\sigma_z\rangle=0$.\\
We can calculate the QFIM using a standard formula for qubits \cite{QFIM}:
\begin{equation}F_{ab}=(\partial_a\vec{r})\cdot(\partial_b\vec{r})+\frac{\vec{r}\cdot(\partial_a\vec{r})\,\,\,\vec{r}\cdot(\partial_b\vec{r})}{1-|\vec{r}|^2}\;, \label{QFIM_bloch}\end{equation}
which expresses the QFI in terms of $\vec{r}$, whose low-coupling behavior we have just obtained. We get an analytic form for $F_{\alpha\alpha}$ by calculating $\partial_\alpha\vec{r}$ and substituting into Eq.\,\ref{QFIM_bloch}:
\begin{equation}F_{\alpha\alpha}\approx\frac{(\partial_\alpha\vec{r})^2}{2}\frac{1}{1-|\vec{r}|^2}=\frac{c^2}{2}\frac{1}{1-(1+\alpha c)^2}\;.\end{equation}
Thus, for $\alpha\rightarrow 0$ the QFI diverges as $F_{\alpha\alpha}\sim -c/(2\alpha)$. A linear fit of $\langle\sigma_x\rangle$ yields $c/2\approx-0.75$, which is just the coefficient obtained by fitting $F_{\alpha\alpha}$. \\
With similar calculations one also finds $F_{\Delta\Delta}\rightarrow0$ and $F_{\alpha\Delta}\rightarrow-\frac{\partial_\Delta c}{2}$.

\section{Static and dynamical \texorpdfstring{$\langle\sigma_z\rangle$}{<sigma z>}}\label{app:sigmaz}
We present the static and dynamical behaviors of $\langle\sigma_z\rangle$, that have been extensively recalled for the interpretation of the results shown in the main text. 

Firstly, in Fig.\,\ref{sigma_z_static}, we plot the GS $\langle\sigma_z\rangle$ as a function of spin-bath coupling $\alpha$ for several values of spin-bath dephasing coupling $\kappa$ at a very small value of the magnetic field $h$. The model parameters are the same as those used in Fig.\,\ref{F_gamma} of the main text. Indeed, under a relatively small dephasing noise, the curve shows a smoothing in the vicinity of the critical point for strong values of spin-bath coupling $\alpha$. These results can be understood by comparing with the plots of panel b of Fig.\,\ref{Static_QFI_new}, considering that the spin-bath coupling $\kappa$ in the Hamiltonian of Eq.\,\eqref{spin-boson-hamiltonian} introduces an effective additional dynamical external field along $z$ axis. With increasing the spin-bath coupling $\kappa$, the smoothing effect becomes more pronounced.

In Fig.\,\ref{sigma_z_tempo}, we show the time evolution of $\langle\sigma_z\rangle$ for the values of $\alpha$ considered in the study of the dynamical QFIM reported in Fig.\,\ref{F_alpha_tempo}. It is evident that the oscillatory behavior of the QFIM is directly related to the coherent evolution of $\langle\sigma_z\rangle$, which, however, disappears as $\alpha \rightarrow 0.5$.

\section{Analysis of dynamical QFIM}\label{app:dynQFIM}
\begin{figure}[htbp]

        \includegraphics[scale=0.2]{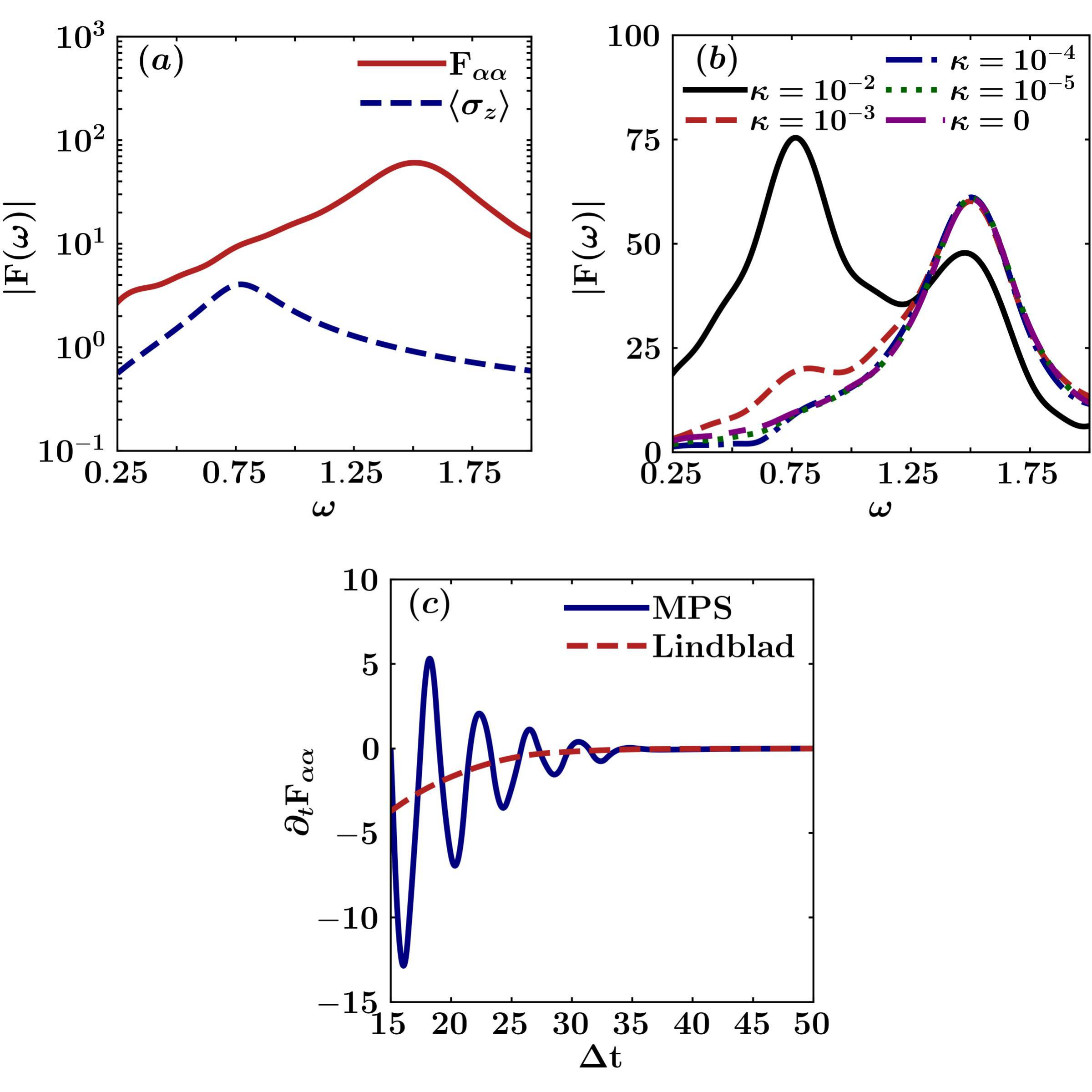}

    \caption{Fourier transform of $F_{\alpha\alpha}$ and $\langle\sigma_z\rangle$ (a) as functions of frequency for the spin-bath coupling $\alpha=0.1$. $\langle\sigma_z\rangle$ exhibits a peak for $\omega=0.75\Delta\approx\Delta_r$, $F_{\alpha\alpha}$ shows a peak close to $\omega=1.5\Delta \approx2\Delta_r$, with $\Delta_r$ renormalized magnetic field. Fourier transforms of $F_{\alpha\alpha}$ as functions of frequency for the spin-bath coupling $\alpha=0.1$ and various $\kappa$ (b). Time derivative of $F_{\alpha\alpha}$ (its flow) as a function of time for $\alpha=0.1$ (b). It is shown starting from $\Delta t=15$ (after the initial transient).}
    \label{FF}
\end{figure}

In panel a of Fig.\,\ref{FF}, we report the Fourier transform of $F_{\alpha\alpha}(t)$ and $\langle\sigma_z\rangle(t)$ for $\alpha=0.1$. 
In the signal of QFIM element, we cut off the long-time tail to the stationary state, responsible for the low-frequency behavior, which is, in any case, not shown in figure. Therefore, the focus is on the peaks associated with the oscillations before equilibration.
The main frequency of $\langle\sigma_z\rangle$ is consistent with the known result for the renormalized magnetic field: $\Delta_r=\Delta(\Delta/\omega_c)^{\frac{\alpha}{1-\alpha}}$ \cite{Hur_2008}. For $F_{\alpha\alpha}$, the dominant frequency is exactly twice that of $\langle\sigma_z\rangle$. This characteristic frequency continues to be present upon the introduction of pure dephasing coupling $\kappa$ (see panel b of Fig.\,\ref{FF}). \\

In the coherent regime of the spin dynamics, the QFI flow, defined as $\partial_t F_{\mu,\nu}$, can be used to characterize non-Markovian effects \cite{Lu_2010,QFIM}. The idea behind the QFI flow is that in a purely Markovian system, information continuously flows out of the system, so the flow remains negative. A positive flow indicates an intake of information, meaning the bath returns part of the information to the subsystem. In panel c of Fig.\,\ref{FF}, we plot the flow of $F_{\alpha\alpha}$ at $\alpha=0.1$ for both MPS and Lindblad solutions. Actually, the Lindblad flow remains negative throughout the dynamics, indicating a one-way information outflow, whereas the MPS solution exhibits an alternating flow, signaling a continuous exchange of information between sub-system and bath.

\section{QFIM in Lindblad approximation}\label{app:lindQFIM}
In the study of dynamical QFIM at zero temperature, we have compared the numerical results with those obtained within the Lindblad approximation. The expression we have used is a particular case of Eq.\,\eqref{Cap1_calcolo_Fisher} of the main text:
\begin{equation}
F_{ab}=2\,\text{Tr}[(\partial_a \rho)(\partial_b \rho)]+\frac{1}{4}\frac{\bigg(\frac{\partial P}{\partial x_a}\bigg)\bigg(\frac{\partial P}{\partial x_b}\bigg)}{det(\rho)}\;,
\label{QFIM}\end{equation}
where $P=\text{Tr}[\rho^2]$. After solving the Lindblad master equation, one obtains the analytical form of $\rho$, which, when substituted into Eq.\,\eqref{QFIM} for $x_a=x_b=\alpha$ gives a closed-form expression for $F_{\alpha\alpha}$:
\begin{align}F_{\alpha\alpha}=&\,4S^2+4t^2
(\gamma^2+(\partial_\alpha\Sigma)^2)|\rho_{01}(0)|^2e^{-2\gamma t} \\
+&\frac{\{S[\rho_{00}(t)-\rho_{11}(t)]-2\gamma t |\rho_{01}(0)|e^{-2\gamma t}\}^2}{\det[\rho(t)]}\;.\end{align}
We have defined
\begin{equation}S(t)\equiv\partial_\alpha\rho_{00}(t)\;,\end{equation}
and $\Sigma$ is the Lamb-Stark shift caused by the spin-bath interaction.
As can be easily seen, $F_{\alpha\alpha}(t\rightarrow+\infty)=0$, since the stationary state is $\ket{\downarrow}$, independently of the coupling. Hence, Lindblad stationary state is not sensitive to variations in $\alpha$: its QFIM vanishes. We remark that the QFIM could be calculated also by means of other methods, such those based on non-equilibrium Green’s functions \cite{perroni}.

\end{document}